\documentclass[12pt,onecolumn,draftclsnofoot]{IEEEtran}
\usepackage{graphicx}
\usepackage[colorlinks,linkcolor=black,anchorcolor=black,citecolor=black]{hyperref}
\usepackage{lineno,hyperref}
\modulolinenumbers[5]
\usepackage[cmex10]{amsmath}
\usepackage{amsfonts}

\usepackage{color}
\usepackage{algorithm}
\usepackage{algorithmic}
\usepackage{array}
\usepackage{multirow}
\usepackage{booktabs}
\usepackage{subfigure}
\usepackage{bm}
\usepackage{graphicx}

\allowdisplaybreaks[4]
\usepackage{amsmath}
\usepackage{underscore}
\usepackage{stfloats}
\usepackage{float}
\usepackage{lipsum}
\usepackage{cite}
\usepackage{amsthm}
\newtheorem{remark}{Remark}

\makeatletter
\newenvironment{breakablealgorithm}
{
	\begin{center}
		\refstepcounter{algorithm}
		\hrule height.8pt depth0pt \kern2pt
		\renewcommand{\caption}[2][\relax]{
			{\raggedright\textbf{\ALG@name~\thealgorithm} ##2\par}%
			\ifx\relax##1\relax 
			\addcontentsline{loa}{algorithm}{\protect\Gammamberline{\thealgorithm}##2}%
			\else 
			\addcontentsline{loa}{algorithm}{\protect\Gammamberline{\thealgorithm}##1}%
			\fi
			\kern2pt\hrule\kern2pt
		}
	}{
		\kern2pt\hrule\relax
	\end{center}
}
\usepackage{amssymb}
\usepackage{amsmath}
\begin{document}

\title{IRS-Enhanced Secure Semantic Communication Networks: Cross-Layer and Context-Awared Resource Allocation}

\author{Lingyi Wang,
Wei Wu,~\IEEEmembership{Member,~IEEE,}
Fuhui Zhou,~\IEEEmembership{Senior Member,~IEEE,}\\
Zhijin Qin,~\IEEEmembership{Senior Member,~IEEE,}
Qihui Wu,~\IEEEmembership{Fellow,~IEEE}

\thanks{This work was supported by the National Key Research and Development Program of China under Grant 2020YFB1807602,
the National Natural Science Foundation of China under Grant 62271267 and the open research fund of National Mobile Communications Research Laboratory, Southeast University under Grant 2024D16.}
\thanks{Lingyi Wang is with the College of Science,
Nanjing University of Posts and Telecommunications, Nanjing, 210003, China
(e-mail: lingyiwang@njupt.edu.cn).}
\thanks{
Wei Wu is with the College of Communication and Information Engineering,
Nanjing University of Posts and Telecommunications, Nanjing, 210003, China,
and also with the National Mobile Communications Research Laboratory, Southeast University, Nanjing, 210096, China
(e-mail: weiwu@njupt.edu.cn).}
\thanks{Fuhui Zhou and Qihui Wu are with the College of Electronic and Information Engineering, 
Nanjing University of Aeronautics and Astronautics, Nanjing, 210000, China
(e-mail: zhoufuhui@ieee.org, wuqihui2014@sina.com).}
\thanks{Zhijin Qin is with Department of Electronic Engineering, Tsinghua University, Beijing, China. 
She is also with the Beijing National Research Center for Information Science and Technology, Beijing, China, and the State Key Laboratory of Space Network and Communications, Beijing, China. (email: qinzhijin@tsinghua.edu.cn).}
}

\maketitle
\begin{abstract}
  Learning-task oriented semantic communication is pivotal in optimizing transmission efficiency by extracting and conveying essential semantics tailored to the specific tasks, such as image reconstruction and classification. 
  Nevertheless, the challenge of eavesdropping poses a formidable threat to semantic privacy due to open nature of wireless communications. 
  In this paper, intelligent reflective surface (IRS)-enhanced secure semantic communication (IRS-SSC) is proposed to guarantee the physical layer security from a task-oriented semantic perspective. 
  Specifically, a multi-layer codebook is exploited to discretize continuous semantic features and describe semantics with different numbers of bits, 
  thereby meeting the need for hierarchical semantic representation and further enhancing the transmission efficiency.
  Novel semantic security metrics, i.e., secure semantic rate (S-SR) and secure semantic spectrum efficiency (S-SSE), are defined to map the task-oriented security requirements at the application layer into the physical layer. 
  To achieve artificial intelligence (AI)-native secure communication, we propose a noise disturbance enhanced hybrid deep reinforcement learning (NdeHDRL)-based resource allocation scheme. 
  This scheme dynamically maximizes the S-SSE by jointly optimizing the bits for semantic representations, reflective coefficients of the IRS, and the subchannel assignment.
  Moreover, we propose a novel semantic context awared state space (SCA-SS) to fusion the high-dimensional semantic space and the observable system state space, which enables the agent to perceive semantic context and solves the dimensional catastrophe problem.
  Simulation results demonstrate the efficiency of our proposed schemes in both enhancing the security performance and the S-SSE compared to several benchmark schemes.
\end{abstract}


\IEEEpeerreviewmaketitle

\section{Introduction}
Task-oriented semantic communication can alleviate network congestion and enhance communication efficiency, which is a promising technology for the sixth-generation (6G) communications \cite{chaccour2022less,luo2022semantic, 9919752}. 
Several works demonstrate the efficiency of semantic communication from mathematical theory \cite{niu2024mathematical,stavrou2023role}.
As one of  emantic communication paradigms,
artificial intelligence (AI)-driven semantic communication can characterize the task-oriented information over semantic symbols by leveraging the formidable knowledge extraction capabilities of machine learning (ML), 
thus significantly reducing the irrelevant information and improving the transmission efficiency \cite{9838470,10122224,10158995}. 
However, the open nature of wireless channel poses the challenge of semantic eavesdropping, leading to serious task exposure and privacy leakage. 
Different from the traditional secure communication that relies on bit-level security at the physical layer, task-oriented secure semantic communication should be concerned with task-level privacy preservation. 
Hence, it is necessary for semantic communication networks to rethink secrecy protection from the task-centered semantic perspective \cite{10123081}.

Research on secure semantic communication mainly focuses on the application layer security \cite{tung2023deep} and the physical layer security \cite{du2023generative}. 
The objective of semantic security at the application layer is to protect the task intention and the task implementation. 
Hence, main research focus is secure semantic coding \cite{10123081} and semantic encryption schemes \cite{10107616} for task-level privacy protection. 
Different from semantic security at the application layer, 
the traditional physical layer technologies for semantic security remain the protection of semantic symbols over the physical channel \cite{wang2023star}. 
Therefore, conventional metrics for symbol protection at the physical layer are insufficient for effectively assessing task-oriented security. 
Similarly, metrics for single intention protection at the application layer struggle to measure the semantic transmission efficiency over physical bits,
which in turn hinders the ability to optimize the physical resource at the semantic level.
To the best knowledge of authors, it remains a tricky problem to measure and ensure the physical layer security from a semantic-level intention protection perspective \cite{yang2023secure, 10292907, du2023generative},
which is essential for efficient secure semantic communications.

Recently, intelligent reflective surface (IRS) has gained significant attention due to its ability to accurately reshape signal propagation in a low-cost and programmable manner \cite{9264659}, 
which has been widely used to improve spectrum efficiency \cite{wang2023intelligent} and communication security \cite{10257607}.  
Specifically, an IRS is a planar array of numerous programmable and passive reflective elements \cite{9847080}. 
Compared to traditional physical security techniques, 
such as relaying and artificial noise \cite{9079457,9197675,9866052}, 
IRS is more energy efficient in manipulating signals by utilizing passive reflection \cite{10257607}. 
The signal reshaping capability of IRS is promising to create a secure transmission environment for semantic communications, 
suppressing semantic eavesdropping and enhancing task performance \cite{10445328}.
Hence, it is crucial to investigate resource allocation with the programmable IRS to bolster semantic security toward future wireless networks.

Deep reinforcement learning (DRL)-based resource allocation schemes exhibit superior real-time performance with intelligence, 
which has been used to rapidly tackle large-scale intricate problems \cite{9653840,wang2023intelligent}. 
Compared to the traditional mathematical optimization methods, such as the block coordinate descent (BCD) \cite{9146170} and the alternating optimization (AO) \cite{9837942},  
DRL-based schemes can manage the network resource in a low-complex and time-efficient way, particularly to address the system complexity caused by the IRS \cite{10257607}.
Moreover, since DRL is AI-native, it works in coordination with ML-driven semantic communication networks \cite{9475174}, where the intelligent agent can perceive the task-related semantics. 
Given the above considerations, DRL-based resource allocation schemes can be a promising solution for the development of efficient and intelligent seucre semantic communication \cite{wang2023adaptive}.

At the application layer, semantic symbol encryption \cite{tung2023deep, 10107616, 10328183} and secure semantic coding \cite{10328183, 10123081} are two main approaches to ensure semantic security.
For the secrecy over encrypted semantic symbols, a public-key cryptographic game was played in \cite{tung2023deep} against the chosen-plaintext attack.
The authors in \cite{10107616} utilized neural networks, trained by adversarial encryption, to embed the encrypted information into the semantics.
However, the public-key encryption scheme \cite{tung2023deep} is computationally expensive, and the encryption embedding scheme \cite{10107616} increases communication costs with extra security information.
For the secure semantic coding, the authors in \cite{10328183} introduced the semantic perturbations to the semantic symbols and misled the semantic interpreter at the eavesdropper with the unseen perturbations.
In \cite{10123081}, further considering the worst case that the semantic interpreters were the same for both the users and eavesdroppers,
the authors directly trained the semantic coding networks with the physical channel differences.
However, the security of semantic coding is threatened since the neural networks are non-interpretable and fragile,
making it hard to be applicable to practical wireless communication environments with dynamic channel.

For the semantic security at the physical layer, the authors in \cite{du2023generative} investigated the covert communication and used a jammer to interfere with the warden to detect semantic information.
However, the artificial jamming requires the extra energy consumption. 
To address this problem, the authors in \cite{wang2023star} applied the IRS to suppress the information leakage to the eavesdropper.
It is worth noting that there have been few works that consider the IRS-enhanced semantic communication networks.
In fact, the research on IRS-assisted conventional communication has demonstrated the significant security gain from the IRS.
The authors in \cite{9733424} utilized the IRS to counteract eavesdropping from the malicious terminals in multiple-input multiple-output (MIMO) systems.
In \cite{10257607} and \cite{wang2023intelligent}, IRS-enhanced secure cognitive networks were considered, where the IRS improved the security of the secondary network and alleviated the interference from the secondary network to the primary network. 
However, similar to the conventional communication, the work \cite{wang2023star} on semantic communication solely aimed to suppress the signal-to-noise ratio (SNR) at the eavesdroppers, i.e., the bit-level security gain,
which had practical limitations from the task perspective. 

Aimed at further enhancing the semantic security efficiency, several semantic resource allocation schemes were investigated.
The authors in \cite{9763856} defined the metric of semantic spectral efficiency, and used the look-up table method to search the optimal semantic symbols assignment. 
In \cite{yan2023qoe}, the authors considered multi-task and multi-cell semantic communications, and utilized deep Q networks to improve the quality-of-experience.
To explore the robust and interpretable resource management, the works \cite{9832831,10016636,10272264} incorporated causal reference into the semantic resource allocation schemes.  
An attention-based DRL method was proposed in \cite{9832831} to optimize the transmitted resource blocks. 
From a gaming perspective, the authors in \cite{10016636} and \cite{10272264} considered resource allocation as a two-player signaling game. 
Extending to the resource allocation for secure semantic communication,
the authors in \cite{du2023generative} applied the diffusion model to generate the optimal resource allocation scheme.
In \cite{wang2023star}, the AO method was used to optimize the transmit power and IRS reflective coefficients.
However, semantic-awared resource allocation remains unexplored since the resource requirements differ with different task-related semantics.  
Moreover, ML-driven semantic communications call for AI-native communication networks with intelligent resource allocation schemes.

Motived by above-mentioned considerations, in this paper, we investigate a context-awared resource allocation scheme for IRS-enhanced secure semantic communication (IRS-SSC) networks.
Since the existing secure communication metrics at the physical layer focus on the bit-level security, firstly, we map the security requirements of the application layer into the physical layer
and define the novel metrics for cross-layer secure semantic communication. Then, the IRS can reshape semantic symbols to suppress the semantic eavesdropping related to the reconstruction task.
To enhance the security performance and robustness of resource allocation, a DRL-based intelligent resource allocation scheme with noise disturbance for action decisions is proposed.
The neuron network-based space fusion method is leveraged to form the semantic-context-aware state space (SCA-SS), which consists of rich semantic space and the observable environment state space.
In this case, our proposed resource allocation scheme can effectively perceive the semantic context and address the high-dimensionality problem, thus enhancing the security efficiency.
The main contributions of this paper are summarized as follows.
\begin{itemize}
  \item Considering the semantic eavesdropping over open wireless communications, 
  we propose a novel IRS-SSC network, 
  where the reflective elements of the IRS can adaptively reshape the propagation of semantic symbols to suppress the semantic eavesdropping and enhance the task performance.
  To discretize the continuous semantic vectors and further enhance the semantic transmission efficiency, 
  a multi-layer codebook is proposed for the first time, 
  where the multi-layer framework aims to trade off the security requirements, semantic difference, task performance and resource limitations.

  \item 
  Since task-oriented semantic security depends on intention protection at the application layer, 
  we map the security requirement of the application layer into the physical layer and redefine cross-layer semantic security (CL-SS) metrics, 
  such as secure semantic rate (S-SR) and secure semantic spectrum efficiency (S-SSE), 
  enabling the physical layer to be aware of intention protection performance.
  We design an adaptive resource allocation scheme achieved by using the noise-disturbance-enhanced hybrid DRL (NdeHDRL). 
  This implementation establishes a robust and AI-native secure semantic communication network.
  Moreover, to effectively address the problem of dimensional catastrophe caused by high-dimensional semantic features and IRS reflective coefficients, 
  we develop a novel SCA-SS to furnish the agent with the semantic context and observable state features. 

  \item Simulation results demonstrate that our proposed NdeHDRL scheme achieves $15.4 \%$ semantic secure performance improvement compared to the HDRL-DS scheme.
  Moreover, our proposed NdeHDRL scheme with SCA-SS obtains $283 \%$ S-SSE improvement compared to the schemes without IRS,
  and $16.5 \%$ S-SSE improvement compared to the NdeHDRL-DS scheme with simple collocation of semantic spaces and observable state spaces (CSO-SS).
  Furthermore, we verify the superior computational efficiency and real-time performance of our proposed scheme compared to the traditional method.
\end{itemize}

The remainder of this paper is organized as follows. 
Section II presents the IRS-SSC network. 
In Section III, the cross-layer security metric is defined, and the formulated resource allocation problem is presented.
Section IV presents the NdeHDRL-based intelligent resource allocation scheme and the SCA-SS.
In Section V, the simulation results and analyses are presented.
Finally, Section VI concludes this paper.

\textit{Notation:} The real matrices set with size $x \times y$ is represented by $\mathbb{R}^{x \times y}$, and
the complex matrices with size $x \times y$ is denoted by $\mathbb{C}^{x \times y}$.
The mathematical expectation is represented by $\mathbb{E}[\cdot]$. 
The element set is denoted by $\{\cdot\}$, and $(x)^+$ means it makes sense only when $x$ is positive.
The function $\arg \underset{x}{\max} g(x)$ denotes the value of $x$ that can maximize $f(x)$.
The $\operatorname{diag}(\mathbf{x})$ denotes a diagonal matrix with diagonal elements corresponding to the elements of $\mathbf{x}$. 
The vector is denoted by $\mathbf{v}$ while the matrix is denoted by $\mathbf{V}$.
The transpose of the vector is denoted by $\mathbf{v}^T$ and the conjugate transpose is denoted by $\mathbf{v}^H$.
The notation \(\left(x\right)^{+}\) denotes the positive part of \(x\). 

\section{IRS-Enhanced Secure Semantic Communication}
\begin{figure*}
  \centering
  \includegraphics[scale=0.65]{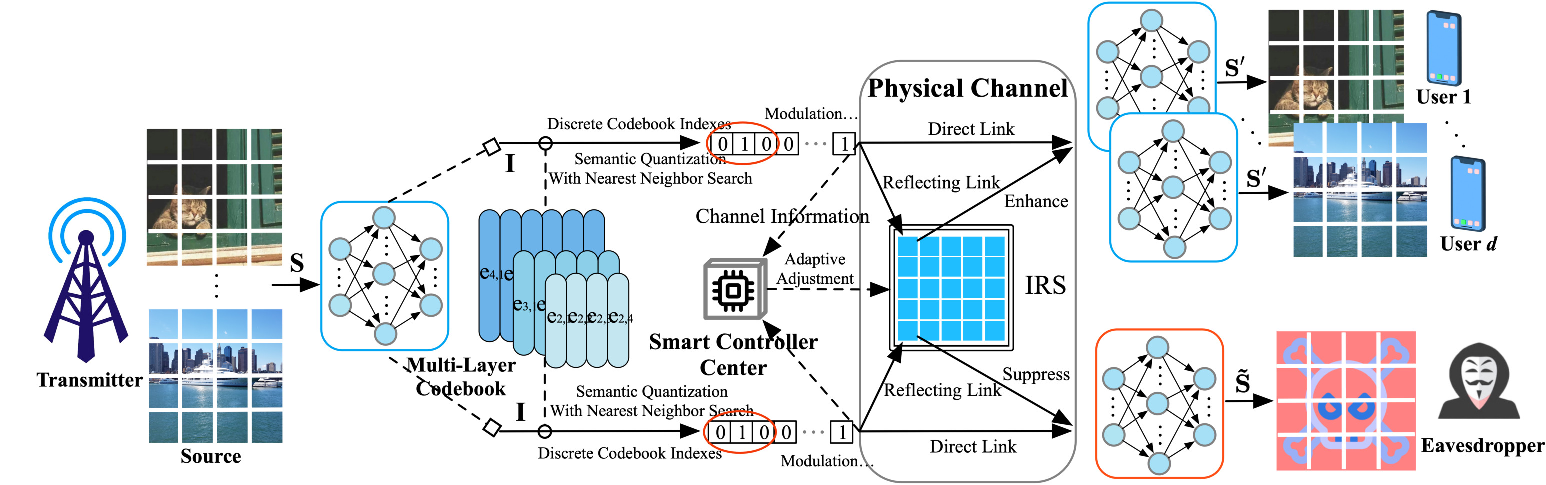}
  \vspace{-0.5cm}
  \caption{An IRS-enhanced secure semantic communication network.}
  \vspace{-0.5cm}
\end{figure*}

\subsection{The Secure Semantic Communication Framework }\label{s2A}
Fig. 1 illustrates an IRS-enhanced downlink semantic communication network with an eavesdropper. 
In this paper, we consider the semantic image reconstruction task.
Since our proposed IRS-enhanced semantic communication framework functions at the physical layer,
this framework can be adapted for all data modalities (e.g. text, voice) and learning tasks (e.g. classification, segmentation). 

Here, the transmitter is equipped with $M$ antennas and a semantic encoder, responsible for capturing the images and encoding the critical information.
The transmitted image $\mathbf{S}$, can be divided into several blocks expressed as $\mathbf{S} = [\mathbf{s}_1, \ldots, \mathbf{s}_i, \ldots, \mathbf{s}_{l \times l}]$, 
where $l \times l$ indicates the number of blocks, and $s_i$ refers to the $i$-th block within the image $\mathbf{S}$.
The image $\mathbf{S}$ is input into the semantic encoder to extract semantic features ${\boldsymbol{\varpi}}$, given by
$
{\boldsymbol{\varpi}}=E_{\omega}(\mathbf{S}),
$
where $E_{\omega}(\cdot)$ is the semantic encoder network with the parameter $\omega$.
The eavesdropper attempts to eavesdrop the semantics ${\boldsymbol{\varpi}}$.
The privacy target is to suppress the semantic eavesdropping and improve the received semantic accuracy at the target receiver.

The codebook serves as the common language to discretize continuous semantic features \cite{10101778} as physical bits,
which can realize the physical layer semantic transmission.
Such common knowledge is learnable based on different application scenarios in order to achieve a high communication gain among legitimate communication participants.
Specifically, the semantic encoder $E_{\omega}(\cdot)$ and decoder $D_{\eta_d}^{-1}(\cdot)$ are pre-trained with the task-related dataset, 
and then the codebooks are trained in the game of the encoder and decoder that are equipped with image reconstruction-relevant semantic knowledge.
The transmitter quantifies the features by searching the most suitable vector-based semantic description $\mathbf{e}_{j}$ in the codebook and then transmits the index of $\mathbf{e}_{j}$.
Here, similar to \cite{10101778}, we adopt the nearest neighbor algorithm to select the description that has the minimum space distance from the target semantic features. 
Hence, the process of describing the semantic features ${\boldsymbol{\varpi}}$ is written as $\mathbf{e} =\arg \min _{\mathbf{e}_{j}}\left\| {\boldsymbol{\varpi}} - \mathbf{e}_{j}\right\|_2, \forall \mathbf{e}_{j}$.

However, we note the differences in semantic richness for each image.
For example, an image with multiple key subjects needs to be described with more semantic details for reconstruction, 
whereas a simple image requires only a small amount of semantics.
Therefore, different images for one specific semantic task can have different levels of semantic representation requirements.
Considering the limited communication resources, 
it is essential to improve the semantic representation efficiency through the characteristics of semantic differences, 
which aims to strike a balance between task performance and resource consumption. 
The balance becomes more complicated when privacy preservation is considered since a larger semantic representation space can increase the semantic confusion degree for eavesdroppers, 
especially in non-ideal eavesdropping environments, where extra attention is paid to semantic security.
With the factors of security requirements, semantic differences, task performance and resource limitations, 
it is hardly possible to obtain an exact selection of codebook layers for a particular task.
Hence, to realize such a multi-layer codebook, we adopt adaptive bits to describe the semantic features in different layers.
In other words, the range of vector indexes, i.e., the size of codebook, is different in each layer.
In case that we adopt $b$ bits for index representations, the common description of ${\boldsymbol{\varpi}}$ is rewritten as
$
  \mathbf{e}_b = \arg \min _{\mathbf{e}_{b,j}}\left\| {\boldsymbol{\varpi}} - \mathbf{e}_{b,j}\right\|_2, \forall \mathbf{e}_{b,j},
$
where $\mathbf{e}_b$ and $\mathbf{e}_{b,j}$ respectively represent the selected semantic description and the $j$-th semantic description within the semantic space represented by $b$ bits. 
The corresponding index of the nearest vectors $\mathbf{e}_b$ serves as the semantic representation, 
and then it is modulated and transmitted over the physical channel.

The receiver is equipped with a single antenna and a semantic decoder, responsible for receiving the semantics and completing the image reconstruction task.
The receiver searches the codebook by using the received index and obtains the recovered semantic description $\mathbf{e}_b^{\prime}$.
It is worth noting that $\mathbf{e}_b^{\prime}$ may not always equal to $\mathbf{e}_b$ due to the potential occurrence of error bits.
The recovered image $\mathbf{S}_d^{\prime}$ at the legitimate user is obtained by
$
  \mathbf{S}_d^{\prime}=D_{\eta_d}^{-1}(\mathbf{e}_b^{\prime}),
$
where $D_{\eta_d}^{-1}(\cdot)$ is the semantic decoder of the legitimate user with the parameters $\eta_d$.
In this paper, we consider the presence of the eavesdropper that eavesdrops semantics.
The eavesdropped image $\hat{\mathbf{S}}_e$ at the eavesdropper can be constructed by
$
  \hat{\mathbf{S}}_e=D_{\eta_e}^{-1}(\hat{\mathbf{e}_b}),
$
where $D_{\eta_e}^{-1}(\cdot)$ is the semantic decoder of the eavesdropper with the parameters $\eta_e$.
Here, we consider a worst-case scenario of eavesdropping, where $\eta_e$ equals $\eta_d$, and the $b$ bits based semantic descriptions can be intercepted by the eavesdropper.

\subsection{IRS-Enhanced Secure Semantic Communications}
In contrast to the current semantic communication research that typically considers an uncontrollable channel environment, 
the programmable IRS can reshape the propagation environment of semantic symbols.
This capability empowers the IRS to establish a perfect communication environment with intelligence, which can facilitate the security of semantic communication.
Note that there are few works that explore the IRS-SSC.
Hence, in this paper, an IRS with $N$ passive reflective elements is employed to improve the system security performance.
The reflective coefficients of the IRS are represented by $\{\alpha_n e^{j \phi_n}\},{n \in N}$,
where $\alpha_n$ and $\phi_n$ represent the amplitude and phase shift of the $n$-th reflective element of the IRS, respectively. 
The diagonal matrix of the reflective coefficients is represented by $\boldsymbol{\Psi}=\mathrm{diag} \{\alpha_1 e^{j \phi_1},\cdots,\alpha_n e^{j \phi_n},\cdots,\alpha_N e^{j \phi_N}\}$.

Let $\mathcal{C}=\{1,\cdots,c,\cdots,C\}$ be the subchannel set and $\mathcal{D} = \{1,\cdots,d,\cdots,D\}$ be the target equipment set.
The subchannel from the transmitter to the legitimate user $d$, from the transmitter to the eavesdropper, from the transmitter to the IRS, from the IRS to the legitimate user $d$ and from the IRS to the eavesdropper
are represented by $\mathbf{h}_{d} \in \mathbb{C}^{M \times 1}$, $\mathbf{w}_{e} \in \mathbb{C}^{M \times 1}$, $\mathbf{h}_{r} \in \mathbb{C}^{N \times 1}$, $\mathbf{g}_d \in \mathbb{C}^{N \times M}$ and $\mathbf{g}_e \in \mathbb{C}^{N \times M}$, respectively.
Note that orthogonal frequency division multiple access (OFDMA) is applied to avoid the interference among multiple users and enhance the spectrum efficiency.
The received signal at the $d$-th legitimate user and the eavesdropped signal at the eavesdropper over the subchannel $c$ are respectively given by 
\begin{subequations}
  \begin{align}
    &y_{d,c} = \underbrace{(\mathbf{h}_{d}^H + \mathbf{h}_{r}^H \boldsymbol{\Psi} \mathbf{g}_d)}_{\mathrm{enhancement}}\mathbf{f}_{d,c} x + n_d, \\
    &y_{e,c} = \underbrace{(\mathbf{w}_{e}^H + \mathbf{h}_{r}^H \boldsymbol{\Psi} \mathbf{g}_e)}_{\mathrm{suppression}}\mathbf{f}_{d,c} x + n_e,
  \end{align}
\end{subequations}
where $\mathbf{f}_{d,c}$ is the transmit beamforming from the transmitter to the user $d$ over the subchannel $c$, $x$ is the target symbol, and $n_d,n_e \sim \mathcal{C N}\left(\mathbf{0}, \sigma^2 \right)$ are the additive white Gaussian noise (AWGN). 
Here, as shown in Fig. 1, the channel state $\mathbf{h}_{d}^H$ and $\mathbf{w}_{e}^H$ are uncontrollable, while the channel again of the legitimate users and the eavesdropper are controllable through adaptive IRS adjustments.

\section{Semantic Performance Metric for Semantic Image Reconstruction}
\subsection{Semantic Coding Metrics}\label{scm}
To access the recovered image at the semantic level, 
we use the structural similarity index measure (SSIM) \cite{wang2004image} and the learned perceptual image patch similarity (LPIPS) \cite{zhang2018unreasonable} as the metric to evaluate the reconstruction similarity.
SSIM measures the differences of luminance, contrast and structural similarity, where a larger SSIM score represents a more similar structure.
LPIPS captures the human visual perception, 
where a larger LPIPS represents more serious visual distortion.
Jointly considering the structure loss captured by SSIM and the human perception loss captured by LPIPS, 
the semantic similarity metric is represented by
\begin{equation}
  \phi(\mathbf{S}, \mathbf{S}_d^{\prime}) = (\operatorname{SSIM} - \xi \operatorname{LPIPS})^{+},
\end{equation}
where $\xi$ is the positive weight coefficient, $\operatorname{SSIM} \in [0,1]$, and $\operatorname{LPIPS} \in [0,1]$.

\subsection{Semantic Security Performance}
The security rate can measure the privacy protection at the physical layer.
The target rate of the $d$-th user over the $c$-th subchannel and the corresponding eavesdropping rate of the eavesdropper are respectively represented by 
\begin{subequations}
  \begin{align}
    {R}_{d,c}= W_c\log_2[1+|(\mathbf{h}_{d}^H + \mathbf{h}_{r}^H \boldsymbol{\Psi} \mathbf{g}_d)\mathbf{f}_{d,c}|^2/{\sigma_d^2}],\\
    {R}_{e,c}= W_c\log_2[1+|(\mathbf{w}_{e}^H + \mathbf{h}_{r}^H \boldsymbol{\Psi} \mathbf{g}_e)\mathbf{f}_{d,c}|^2/{\sigma_e^2}],
  \end{align}
\end{subequations}
where $W_c$ is the bandwidth of the $c$-th subchannel.
Hence, the achievable secrecy rate of the $d$-th user over the $c$-th subchannel can be obtained by
\begin{equation}
R^s_{d,c}=\left({R}_{d,c}-{R}_{e,c}\right)^{+}.
\end{equation}

Unlike conventional secure communication that primarily emphasizes the physical layer security at the bit level, 
task-oriented semantic communication requires a redefinition of security performance from a task-level security perspective at the application layer. 
The work in \cite{10123081} has provided an example of a secure image reconstruction task,
where the distance between the eavesdropped image $\hat{\mathbf{S}}_e$ and all black image $\mathbf{0}$ is considered as the eavesdropping gain.
However, this metric fails to incorporate human perception loss and semantic distortion. 
Notably, due to the semantic noise during transmission and IRS-enhanced signal suppression,
the eavesdropped semantics may be deviated even though it contains rich color information compared to an all-black image.
In such a case, it can be considered as secure semantic transmission. 
Moreover, the human perception loss is able to evaluate the human visual difference between the original image and the eavesdropped image, which can be added to the eavesdropping gain through our proposed $\phi$.
Hence, we introduce a novel privacy metric $\phi^s$ for secure semantic communication.
The achievable semantic security performance at the target equipment $d$ over the $c$-th subchannel can be represented by
\begin{equation}
  \phi^s_{d,c}= (\phi(\mathbf{S}, \mathbf{S}_d^{\prime}) - \epsilon^s_{d,c} \phi(\mathbf{S}, \hat{\mathbf{S}}_e))^{+},
\end{equation}
where $\epsilon^s_{d,c}$ is the effectiveness of eavesdropping. 
If $\phi(\mathbf{S}, \mathbf{S}_d^{\prime})-\phi(\mathbf{S}, \hat{\mathbf{S}}_e) \leq \phi^{th}_{d,c}$, $\epsilon^s_{d,c}=0$; otherwise, $\epsilon^s_{d,c}=1$. 

\subsection{Cross-Layer Semantic Security Metrics}
Existing semantic security metrics independently assess the physical layer security \cite{wang2023star} or the application layer security \cite{10123081}.
It is the first work to consider the CL-SS metrics, secure semantic rate (S-SR) and secure semantic spectrum efficiency (S-SSE), where the security requirement of the task-related application layer is mapped into the physical layer.
The S-SR is given by
\begin{equation}
  \widetilde{R}^{s}_{d,c} = \frac{\phi^s_{d,c}}{|{\boldsymbol{\varpi}}|} \times \frac{{R}^{s}_{d,c}}{b},
\end{equation}
where $|{\boldsymbol{\varpi}}|$ is the number of the transmitted semantic symbols ${\boldsymbol{\varpi}}$.
To the left of the multiplication operation, 
the ratio of the task-related security performance to the number of semantic symbols indicates the semantic symbol security efficiency in the unit of semantic unit ($sut$). 
To the right of the multiplication operation, 
the ratio of the physical secure transmission rate to the number of bits per semantic symbol indicates the physical semantic secure transmission efficiency in the unit of per second ($/s$).
In this context, ${1}/|{\boldsymbol{\varpi}}|$ is regarded as the semantic unit ($sut$), and the unit of the $\widetilde{R}^{s}_{d,c}$ is $sut/s$.
Note that the semantic unit is independent of bit symbols, it is clear that semantic communications prioritize the achievable semantic similarity at the application layer rather than the accurate physical bits,
which emphasizes the necessity of CL-SS metrics.
\begin{remark}
  The $|{\boldsymbol{\varpi}}|$ determines the carried semantic information for each semantic unit, as a larger number can accommodate more detailed semantic information for tasks.
  The optimal choice of $|{\boldsymbol{\varpi}}|$ can be predetermined, as it remains relatively independent from the physical layer transmission while pertaining to the task.
  Hence, the $1/|{\boldsymbol{\varpi}}|$ can be the basis unit since it is the certain value for a specific semantic communication system.
  In contrast, the number of bits $b$ for transmission makes sense to both packet size at the physical layer and task performance at the application layer.
  Since more bits can support a larger semantic representation space,
  it is meaningful and practical to research the adaptive bits for wireless semantic communications.
\end{remark}

Furthermore, the S-SSE is given by
\begin{equation}
  \Phi^{s}_{d,c} = \frac{\widetilde{R}^{s}_{d,c}}{W_c}.
\end{equation}
It is noteworthy that our proposed S-SSE is applicable to any codebook-enabled digital semantic communication networks.
The unit of the S-SSE $\Phi^{s}_{d,c}$ is $sut/s/\mathrm{Hz}$.

\subsection{Resource Allocation Problem Formulation}
Let $\rho_{d, c}$ be whether the $d$-th UE occupies the $c$-th subchannel or not.
If the $c$-th subchannel is occupied by the $d$-th UE, $\rho_{d, c}=1$; otherwise $\rho_{d, c}=0$.
Hence, the S-SSE of semantic communication networks can be represented by 
\begin{equation}
  \Upsilon=\sum_{d=1}^D \sum_{c=1}^C \rho_{d, c}\Phi^s_{d,c}.
\end{equation}

To maximize the overall S-SSE, the bits for semantic representations, the reflection coefficients of the IRS and the subchannel assignment are jointly optimized.
Let $\mathbf{B}=\{b_d\}_{\forall d \in \mathcal{D}}$ and $\boldsymbol{\rho} = \{\{\rho_{d, c}\}_{{\forall c \in \mathcal{C}}}\}_{\forall d \in \mathcal{D}}$ be the bit decision set and subchannel assignment set, respectively.
The optimization problem is formulated by
\begin{subequations}\label{op}
  \begin{align}
   \mathbf{P}: &\max_{\mathbf{B},\boldsymbol{\Psi},\boldsymbol{\rho}} \Upsilon \\ 
   \text{ s.t. }  
   & b_d \in \{b_{\mathrm{min}},b_{\mathrm{min}}+1,\cdots,b_{\mathrm{max}}\}, \forall d,\\
   &\phi_n \in\{0,2 \pi\}, \forall n,\\
   &0 \le \left|[\boldsymbol{\Psi}]_{n,n}\right|\le 1,\forall n\\
   &\rho_{d, c} \in\{0,1\}, \forall d, \forall c,\\
   &\sum_{c=1}^C\rho_{d, c}=1, \forall d,\\
   &\sum_{d=1}^D\rho_{d, c}=1, \forall c,
  \end{align}
\end{subequations}
where $b_{\mathrm{min}}$ and $b_{\mathrm{max}}$ respectively represent the minimum and maximum bits for the semantic vector indexes.
Constraint (\ref*{op}b) limits the size of each semantic space in the codebook.
Constraints (\ref*{op}c) and (\ref*{op}d) stipulate the permissible ranges for IRS reflective coefficients.
Constraints (\ref*{op}e), (\ref*{op}f) and (\ref*{op}g) pertain to the subchannel assignment, 
ensuring that each subchannel is occupied by at most one UE, and each UE can occupy only one subchannel.

\section{The Proposed Intelligent Semantic Resource Allocation Scheme}
This section presents our designed intelligent resource allocation scheme for IRS-SSC networks illustrated in Fig. 2.
The training process of our proposed scheme consists of three stages.
In the first stage, we pre-train the semantic coding and multi-layer codebook.
In the second stage, the S-SSE maximization problem (\ref*{op}) is solved by using Markov Decision Process (MDP). 
Based on the defined MDP model, we propose NdeHDRL for robust and intelligent resource allocation.
A novel SCA-SS is designed to provide the agent with semantic context and address the dimensional catastrophe caused by the high-dimensional semantic space and observable state space.
In the third stage, we fine-tune the entire resource allocation scheme.
\begin{figure*}
  \centering
  \includegraphics[scale=0.66]{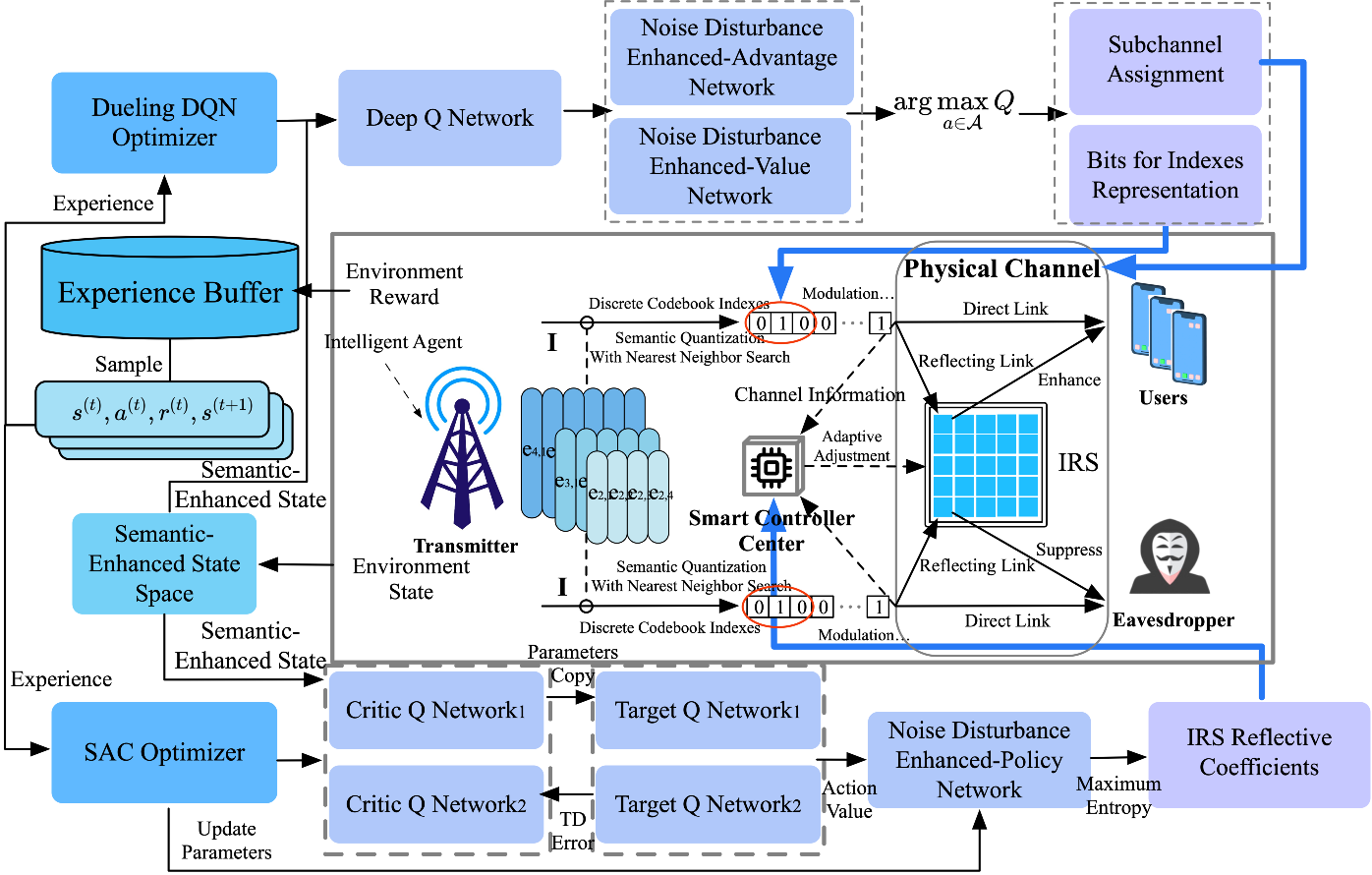}
  \caption{Our proposed intelligent resource allocation scheme for IRS-SSC networks.}
  \vspace{-0.5cm}
\end{figure*}

\subsection{Pre-Training Semantic Coding and Multi-Layer Codebook}
Based on the semantic similarity defined in Section \ref{scm}, the loss function for semantic coding is expressed as
\begin{equation}\label{scm1}
  \mathcal{L}_{\phi}(\mathbf{\omega},\mathbf{\eta}_d) = 1-\operatorname{SSIM} + \xi \operatorname{LPIPS},
\end{equation}
where the structure loss and human perception loss are jointly considered to assess the image reconstruction task.
The semantic coding networks are independently trained to minimize Eq. (\ref{scm1}).
Then, the multi-layer codebook is trained with semantic codebook networks to represent the semantic space and discretize the continuous semantic symbols. 
Specifically, the total loss of adopting $b$ bits for semantic representations is represented by
\begin{equation}
  \begin{split}
  \mathcal{L}_c\left(b;\mathbf{\omega},\mathbf{\eta}_d\right)= & \mathcal{L}_{\phi} + \beta_c\left\|\operatorname{tg}\left[ {\boldsymbol{\varpi}}\right]-\mathbf{e}_b\right\|_2^2 \\ &+\beta_e\left\| {\boldsymbol{\varpi}}-  \operatorname{tg}\left[\mathbf{e}_b\right]\right\|_2^2 - \beta_e \left\|\mathbf{E}_b^T \mathbf{E}_b\right\|_2,
\end{split}
\end{equation}
where $\operatorname{tg}$ stands for the truncated gradient, i.e., the gradient backpropagation stops passing after the truncation.
The item $\left\|\operatorname{tg}\left[ {\boldsymbol{\varpi}}\right]-\mathbf{e}_b\right\|_2^2 +\beta_c\left\| {\boldsymbol{\varpi}}-\operatorname{tg}\left[\mathbf{e}_b\right]\right\|_2^2$ is the codebook loss, while the item $\left\| {\boldsymbol{\varpi}}-\operatorname{tg}\left[\mathbf{e}_b\right]\right\|_2^2$ is the coding loss. 
To expand the semantic representation space, the item $\left\|\mathbf{E}_b^T \mathbf{E}_b\right\|_2$ is considered to maximize the semantic distance between vectors in the codebook, i.e., the semantic vectors are orthogonal to each other.
Specifically, the normalized basis vectors are represented by
\begin{equation}
  \mathbf{E}_b \triangleq \left[\frac{\mathbf{e}_{b,1}}{\left\|\mathbf{e}_{b,1}\right\|}, \frac{\mathbf{e}_{b,2}}{\left\|\mathbf{e}_{b,2}\right\|}, \cdots, \frac{\mathbf{e}_{b,L_b}}{\left\|\mathbf{e}_{b,L_b}\right\|}\right],
\end{equation}
where $L_b$ is the maximum integer that can be represented by $b$ bits. 
Details regarding the training of semantic coding and the multi-layer codebook are shown in \textbf{Algorithm 1}.

\subsection{Semantic-Awared MDP Problem}
MDP forms the foundational basis for designing the RL model of our proposed IRS-SSC network. 
Hence, the S-SSE maximization problem formulated in (\ref{op}) is a MDP problem. 
In this context, the IRS-SSC network serves as the interactive environment, and
the control center of the BS is designed as the intelligent agent, connecting the smart controller of the IRS.
The following are the key elements of the MDP problem.

The state space describes the environment state of the intelligent agent.
However, there are two challenges for constructing a state space of the IRS-SSC networks.
\begin{itemize}
\item Challenge 1: How does the intelligent agent perceive semantic context efficiently?
\item Challenge 2: How to deal with the ultra-high-dimensional state space with semantics?
\end{itemize}

For Challenge 1, as analyzed in Section \ref*{s2A}, different images have different amounts of semantic information.
Hence, it is important for the agent to understand this criterion, i.e., semantic context.
For Challenge 2, the state space consists of an observable state space and a task-related semantic space. 
The semantic features in the semantic space are high-dimensional while the dimensionality of the observable state space increases with the number of the IRS reflective elements.
Hence, the state space is ultra-high-dimensional, especially with a large number of IRS elements.
To overcome these two critical challenges, we propose a novel SCA-SS. 
The details are depicted in Fig. 3.
\begin{figure*}
  \centering
  \includegraphics[scale=0.7]{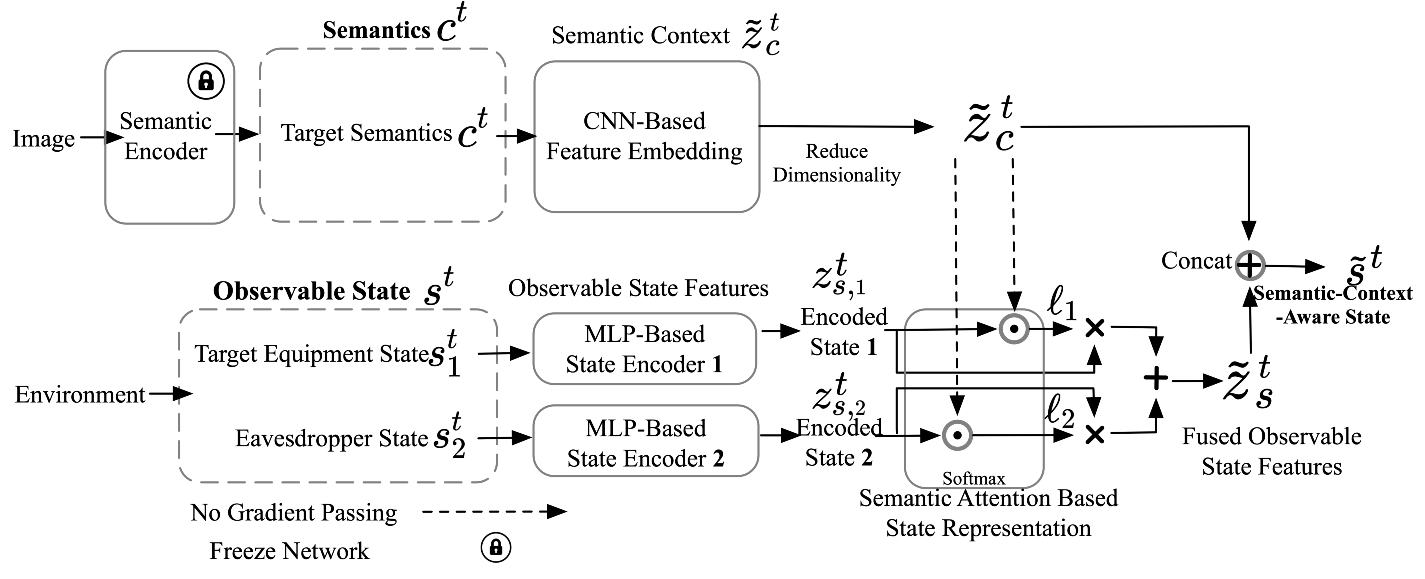}
  \vspace{-0.5cm}
  \caption{The semantic context awared state for the MDP-based semantic resource allocation problem.}
  \vspace{-0.5cm}
\end{figure*}

Let $\mathcal{S}$ be the observable state space, which consists of the channel information of the users and the eavesdropper, and the previously selected actions,
and the superscript $t$ be the timestamp of the communication system.
The observable state $s^{(t)} \in \mathcal{S}$ is represented by $s^{(t)}=\{a^{(t-1)},\{\mathbf{h}_d^{(t)}\},\{\mathbf{h}_e^{(t)}\}\}$, 
which can be divided into two parts, $s^t_1=\{a^{(t-1)},\{\mathbf{h}_d^{(t)}\}\}$ and $s^t_2=\{a^{(t-1)},\{\mathbf{h}_e^{(t)}\}\}$.
Here, the action of the previous step $a^{(t-1)}$ is included in the current state to provide additional information about the dynamics of the environment, 
which aims to better approximate Markovianity, capture temporal dependencies, and understand the agent behavior.
Let $\mathcal{C}$ be the semantic space, which consists of the semantic context.
The semantic features $c^{(t)} \in \mathcal{C}$ are represented by $c^{(t)}=\{\{\mathbf{I_d}^{(t)}\}\}$.
Two separate MLP-based encoders ${E_{\tau_{j}}}(\cdot),j \in \{1,2\}$ are utilized to extract the features of the observable state, where the state features are respectively represented by $z_{s,1}^{(t)} = E_{\tau_1}(s^{(t)}_1)$ and $z_{s,2}^{(t)} = E_{\tau_{2}}(s^{(t)}_2)$. 
Meanwhile, a CNN-based embedding encoder is used to extract the semantic context from the semantic features, where the semantic context is represented by $\tilde{z}_{c}^{(t)}=E_{\kappa}(c^{(t)})$.
The observable state features with the semantic context are obtained by using the Hadamard product $\odot$, which is represented by
\begin{equation}
  \hat{z}_{s,j}^{(t)} = z_{s,j}^{(t)} \odot \tilde{z}_{c}^{(t)}, \forall j \in \{1,2\}.
\end{equation}
\begin{breakablealgorithm}
  \caption{Pre-Training Semantic Coding Network}
  \begin{algorithmic}[1]
    \STATE Initialize the network parameters $\boldsymbol{\omega}$, $\boldsymbol{\eta}$ and the codebook $\{\{\mathbf{e}_{b,j}\}_b\}$;
    \FOR{Epoch $p=1$ $\rightarrow$ $P$}
    \STATE Random $b$;
    \STATE \textbf{Transmitter:}
    \STATE \par \hspace{1cm} Batch images $\{\mathbf{S}\}$ $\rightarrow$ $\mathbf{S}$;
    \STATE \par \hspace{1cm} $E_{\boldsymbol{\omega}^*}(\mathbf{S})$ $\rightarrow$ ${\boldsymbol{\varpi}}$;
    \STATE \par \hspace{1cm} $\mathbf{e}_{b,j}({\boldsymbol{\varpi}})$ $\rightarrow$ $\mathbf{e}_b$;
    \STATE \textbf{Receiver:}
    \STATE \par \hspace{1cm} Receive $\mathbf{e}_{b}$;
    \STATE \par \hspace{1cm} $D_{\eta}^{-1}(\mathbf{e}_{b})$ $\rightarrow$ $\mathbf{S}_d^{\prime}$;
    \ENDFOR
    \RETURN Pre-trained semantic coding network: $E_{\boldsymbol{\omega}^*}(\cdot),D_{\boldsymbol{\eta}^*}^{-1}(\cdot),\{\{\mathbf{e}_{b,j}^*\}_b\}$.
  \end{algorithmic}
\end{breakablealgorithm}
Here,
the Hadamard product enables element-by-element interactions between the observable state space and the semantic context, 
highlighting the attention of states in a particular semantic background and guiding the agent to capture finer-grained semantic relationships.
The attention over the observable state features of the user and eavesdropper is obtained by using the operation $softmax(\cdot)$, 
then the fused observable state features is expressed as
\begin{equation}
    \tilde{z}_{s}^{(t)} = \ell_1 \times z_{s,1}^{(t)} + \ell_2 \times z_{s,2}^{(t)},
\end{equation}
where $\times$ represents the element-wise multiplication.
The semantic-context-aware state is obtained by combining the fused state features and semantic context, which is expressed as
\begin{equation}
  \tilde{s}^{(t)} = \tilde{z}_{s}^{(t)} \oplus \tilde{z}_{c}^{(t)},
\end{equation}
where $\oplus$ is the concat operation. 
\begin{remark}
  Here, $|c^{(t)}| \ll |\tilde{z}_{c}^{(t)}|$ and $|s^{(t)}| \ll |\tilde{z}_{s}^{(t)}|$. 
  Moreover, the dimensionality $|s^t|$ is independent of the number of IRS reflective elements $N$.
  Hence, our proposed scheme is less likely to fall into a local optimal solution even when $N$ is ultra-large.
\end{remark}

Let SCA-SS be $\tilde{\mathcal{S}}$, composed of $\{\tilde{s}^{(t)}\}$.
The state transition probability is related to the probability distribution of the semantic context $p(c^{(t)}\mid\mathbf{S})$,
which is represented by $\mathcal{T}(s^{(t+1)} \mid s^{(t)}, a^{(t)}, p(c^{(t)}\mid\mathbf{S}))$.
The details of the SCA-SS construction are given in \textbf{Algorithm 2}.
\begin{breakablealgorithm}
  \caption{The SCA-SS Construction}
  \begin{algorithmic}[1]
    \STATE Give the parameters of CNN $\kappa$ and the state encoders $\tau_j$, $j=\{1,2\}$;
    \STATE $E_{\tau_j}(s^{(t)}_j)$ $\rightarrow$ $z_{s,j}^{(t)}$;
    \STATE $E_{\kappa}(c^{(t)})$ $\rightarrow$ $\tilde{z}_{c}^{(t)}$;
    \STATE $z_{s,j}^{(t)} \odot z_{c}^{(t)}$ $\rightarrow$ $\hat{z}_{s,j}^{(t)}$;
    \STATE Attention $softmax(\hat{z}_{s,j}^{(t)})$ $\rightarrow$ $\ell_j$;
    \STATE Fusion $\sum_{j=1}^{2} \ell_j z_{s,j}^{(t)}$ $\rightarrow$ $\tilde{z}_{s}^{(t)}$;
    \STATE $\tilde{z}_{s}^{(t)} \oplus \tilde{z}_{c}^{(t)}$ $\rightarrow$ $\tilde{s}^{(t)}$;
    \STATE \textbf{Return} semantic context awared state $\tilde{s}^{(t)}$.
  \end{algorithmic}
\end{breakablealgorithm}

The action space of the IRS-SSC network is represented by $\mathcal{A}$.
The action $a^{(t)} \in \mathcal{A}$ involves the bit number for index representations in the codebook $\mathbf{B}^{(t)}$, reflective coefficients of the IRS $\boldsymbol{\Psi}^{(t)}$ and the subchannel assignment $\boldsymbol{\rho}^{(t)}$. 
At the timestamp $t$, the action $a^{(t)}$ can be represented by
$
a^{(t)}=\{\mathbf{B}^{(t)},\boldsymbol{\Psi}^{(t)},\boldsymbol{\rho}^{(t)}\}.
$

The reward function is designed to encourage the intelligent agent to the maximize the S-SSE $\Upsilon$.
Since the quality of semantic symbols is strongly correlated with task performance,
channel gains are added as penalty items to guide the agent to enhance the target semantic symbols at the legitimate users and suppress the eavesdropped symbols at the eavesdroppers.
Specifically, the channel gains of the users and the eavesdropper are represented by $\upsilon_d=\|\mathbf{h}_{d}^H + \mathbf{h}_{r}^H \boldsymbol{\Psi} \mathbf{g}_d\|$ and $\upsilon_e=\|\mathbf{w}_{e}^H + \mathbf{h}_{r}^H \boldsymbol{\Psi} \mathbf{g}_e\|$.
With two punishment items, the reward function is written as
$r^{(t)}= \Upsilon^{(t)} + w(\upsilon_d-\upsilon_e)$, where $w$ is a given weight coefficient.
Over an extended and stable transmission period, the distribution of the transmitted images $\mathrm{Pr}(\mathbf{S})$ can be considered to be virtually unchanged.
Hence, without prejudice to generality, the state transfer matrix $\mathbf{T}$ is randomly set as assess the data source, $\mathbf{T} \sim \mathrm{Pr}(\mathbf{S})$, while the wireless environment is dynamically changed.

The policy adopted by our proposed intelligent agent is represented by $\pi$, 
where the target of the training is to find the optimal policy $\pi^*$.
When employing the policy $\pi$, the Q value associated with a decision pair $\langle s,a \rangle$ is represented by
$
  Q_\pi(s, a)=\mathbb{E}_{\pi}(\sum_{t=1}^{\infty} \gamma^{(t-1)} r^{(t)} \mid s^{(t)}=s, a^{(t)}=a),
$
where $\gamma$ is the discount coefficient to evaluate the future rewards.
The value of the state $s$ can be evaluated by
$
  V_\pi(s) =\mathbb{E}_{\pi}(\sum_{t=1}^{\infty} \gamma^{(t-1)} r^{(t)} \mid s^{(t)}=s)= \sum_{a \in A} \pi(a \mid s) Q_\pi(s, a).
$
Hence, the advantage function is represented by
$
  A_\pi(s, a)=V_\pi(s)-Q_\pi(s, a),
$
and the optimal policy $\pi^*$ is written as
$
  \pi^* = \max_{\pi} \mathbb{E}_{\pi}(\sum_{t=1}^{\infty} \gamma^{(t-1)} r^{(t)} \mid s^{(0)})= \max_{\pi} V_\pi(s^{(0)}).
$

\subsection{Intelligent Resource Allocation Scheme Based on NdeHDRL}
Since the intelligent agent explores the dynamic wireless communication environment,
there is a demand to enhance the robust performance and exploratory capacity of DRL in the SCA-SS.
To meet this demand, we propose a noise-disturbance-enhanced hybrid resource allocation scheme, named NdeHDRL.
Specifically, we introduce the NoisyNet \cite{fortunato2017noisy}, where the noise disturbance is used to improve action decisions.
Based on the NdeHDRL design, our proposed NdeHDRL-Dueling DQN-Soft Actor-Critic (SAC) (NdeHDRL-DS) is also presented as follows.

\subsubsection{Factorised Gaussian Noise Based Perturbation}
The noise disturbance enhanced network is written as
$
  \mu \triangleq  \theta + \zeta \odot \varepsilon,
$
where $\theta$ and $\zeta$ are the learnable network parameters, and the $\varepsilon$ is the noise perturbation.
Aiming to improve the generalization capabilities and adaptively control of weight perturbations,
the factorised Gaussian noise is applied to serve as perturbations, written as
\begin{subequations}
  \begin{align}
  \varepsilon_{i, j}^w & =f\left(\varepsilon_i\right) f\left(\varepsilon_j\right), \\
  \varepsilon_j^b & =f\left(\varepsilon_j\right),
  \end{align}
\end{subequations}
where $f(x)$ is the real-valued function, and $\varepsilon_i$ and $\varepsilon_j$ represent different nosie components.
Similar to \cite{fortunato2017noisy}, we set $f(x)=\operatorname{sgn}(x) \sqrt{|x|}$, $\theta_{i,j} \sim \mathcal{U}\left[-\frac{1}{\sqrt{p}},+\frac{1}{\sqrt{p}}\right]$ and $\zeta_{i,j} \sim \frac{\theta_{0,0}}{\sqrt{p}}$.
The loss function for $\theta$ and $\zeta$ can be expressed as
$
\nabla \mathcal{L}_{\theta, \zeta}=\nabla \mathbb{E}[L(\mu)]=\mathbb{E}\left[\nabla_{\theta, \zeta} L(\theta + \zeta \odot \varepsilon)\right].
$
Further reducing the computation complexity, Monte Carlo approximation of the gradient is applied, where the single-step optimization is rewritten as
\begin{equation}
\nabla \mathcal{L}_{\theta, \zeta}\approx \nabla_{\theta, \zeta} L(\theta + \zeta \odot \varepsilon).
\end{equation}
By utilizing the factorized Gaussian noise based perturbation, we can adapt the agent to the time-varying wireless communication environments and rich semantic context, as shown in \ref*{s-c}.

\subsubsection{NdeHDRL-Dueling DQN for Adaptive Bit Allocation and Subchannel Assignment}
NoisyNet-Dueling DQN exhibits superior exploration performance in the discrete space compared to the Dueling DQN \cite{fortunato2017noisy}. 
By adaptive adjustments of the weighted noise, NoisyNet-Dueling DQN effectively explores the high-dimensional semantic-enhanced state space,
ultimately achieving the optimal bit allocation for vector representations and the subchannel assignment.

Let $\theta$ and $\zeta$ be the parameters of the eval networks, $\theta^{-}$ and $\zeta^{-}$ be the parameters of the target newtorks in the NoisyNet-Dueling DQN.
To enhance the algorithm stability and solve the Q overestimation problem,
the Q value in NoisyNet-Dueling DQN is obtained by
$
  Q(\tilde{s}^{(t)}, a^{(t)}; \theta, \zeta)=V(\tilde{s}^{(t)}; \theta, \zeta)+(A(\tilde{s}^{(t)}, a^{(t)}; \theta, \zeta)-\frac{1}{|\mathcal{A}|} \sum_{a} A(\tilde{s}^{(t)}, a; \theta, \zeta)),
$
where $|\mathcal{A}|$ represents the dimension of the $\mathcal{A}$.
The tuple $\{\tilde{s}^{(t)}, a^{(t)}, r^{(t)}, \tilde{s}^{(t+1)}\}$ is sampled from the experience buffer $\mathcal{T}$, 
then the loss function of NdeHDRL-Dueling DQN training is calculated by
$
\mathcal{L}_{\kappa^*,\tau_{j}^*}(\theta, \zeta) =\mathbb{E}_{(\tilde{s}^{(t)}, a^{(t)}, r^{(t)}, \tilde{s}^{(t+1)}) \sim \mathcal{T}}[r^{(t)}+\gamma Q(\tilde{s}^{(t+1)}, a^{*(t+1)}(\tilde{s}^{(t)}), \varepsilon^{\prime} ; \theta^{-}, \zeta^{-})-Q(\tilde{s}^{(t)}, a^{(t)}, \varepsilon ; \theta, \zeta)]^2.
$
Here, the optimal action $a^{*(t+1)}$ is obtained with noise disturbance, represented by
\begin{equation}
  a^{*(t+1)}(\tilde{s}^{(t)}; \theta, \zeta)=\arg \max _{a \in \mathcal{A}} Q\left(\tilde{s}^{(t)}, a, \varepsilon^{\prime} ; \theta, \zeta\right).
\end{equation}

\subsubsection{NdeHDRL-SAC for IRS Reflective Coefficients Adjustment}
The stochastic strategy-based SAC is considered as one of the state-of-the-art (SOTA) algorithms to address the continuous action space.
The maximum entropy policy in SAC can encourage the strategic diversity, which is beneficial to dealing with high-dimensional IRS reflective coefficients.
Hence, NdeHDRL-SAC is proposed in this paper to achieve powerful exploration capabilities and robust algorithm performance.

The entropy regularization Q value is rewritten as
$
Q_\pi(\tilde{s}^{(t)}, a^{(t)}) = r^{(t)}+\gamma(Q_\pi(\tilde{s}^{(t+1)}, \tilde{a}^{(t+1)})$
$ - \alpha \log \pi(\tilde{a}^{(t+1)} \mid \tilde{s}^{(t+1)})).
$
Here, $\tilde{a}^{(t+1)} \sim \pi(\cdot \mid \tilde{s}^{(t+1)})$ is the resample trick.
The maximum entropy-enhanced optimal policy is represented by
$
\pi^*(\tilde{s}^{(t)}, a^{(t)})=\arg \underset{\pi}{\operatorname{\max}} \, \mathbb{E}_\pi[(\sum r(\tilde{s}^{(t)}, a^{(t)}))+\nu H(\pi(\cdot \mid \tilde{s}^{(t)}))],
$
where $H(\pi(\cdot \mid \tilde{s}))$ is the entropy term and $\nu$ is the constant coefficient.
The network parameter of the eval Q networks is $\delta_j$, and the network parameter of the target Q networks is $\delta^{-}_j$, ${j=1,2}$.
The loss function of the Q-networks in SAC is represented by 
$
\mathcal{L}_{{\kappa^*,\tau_{j}^*}}(\delta_j)= \mathbb{E}_{(\tilde{s}^{(t)}, a^{(t)}, r^{(t)}, \tilde{s}^{(t+1)}) \sim \mathcal{T}}[(y(\tilde{s}^{(t+1)}, a^{(t+1)};\delta_j)-Q(\tilde{s}^{(t)}, a^{(t)};\delta_j))^2].
$
The target value $y$ is calculated by
$
y(\tilde{s}^{(t+1)}, a^{(t+1)};\delta_j)=r^{(t)}+\gamma \min _{j\in\{1,2\}} Q_{\delta_j^{-}}(\tilde{s}^{(t+1)}, a^{(t+1)})-\Lambda \log \pi(a^{(t+1)} \mid \tilde{s}^{(t+1)}),
$
where $\Lambda$ is the given entropy temperature.

The parameters of the policy network and the target policy network are $\lambda$ and $\lambda^{-}$, respectively.
The noise-disturbance-enhanced policy network is expressed by
\begin{equation}
\tilde{a}^{(t+1)}(\tilde{s}^{(t)}; \lambda, \vartheta)=\tanh (\mu(\tilde{s}^{(t)};\lambda)+\sigma(\tilde{s}^{(t)};\lambda) \odot \vartheta),
\end{equation}
where $\vartheta$ is the factorised Gaussian noise in NdeHDRL-SAC.
The loss function of the policy is expressed by
$\mathcal{L}_{\kappa^*,\tau_{j}^*}(\lambda,\vartheta)=\mathbb{E}_{(\tilde{s}^{(t)}, a^{(t)}, r^{(t)}, \tilde{s}^{(t+1)}) \sim \mathcal{T}}[\Lambda \log \pi(a^{(t)} \mid \tilde{s}^{(t)};\lambda,\vartheta)-\min _{j=1,2} Q(\tilde{s}^{(t)}, a^{(t)};\delta_j)].$

When training the semantic-context-enhanced network,
the network parameters of the resource allocation scheme are fixed, and the loss function is represented by
\begin{equation}
  \mathcal{L}_{\lambda^*,\vartheta^*,\theta^*, \zeta^*}(\kappa,\tau_{j})=\mathcal{L}_{\lambda^*,\vartheta^*} + \mathcal{L}_{\theta^*, \zeta^*}.
\end{equation}

The training details of our proposed semantic context awared intelligent resource allocation scheme by using NdeHDRL-DS for IRS-SSC is given in \textbf{Algorithm 3}.
\vspace{0.4cm}
\begin{breakablealgorithm}
  \caption{Our Proposed Semantic-Contex-Aware Scheme Using NdeHDRL-DS}
  \begin{algorithmic}[1]
    \STATE Obtain the pre-trained semantic coding networks based on \textbf{Algorithm 1}: $\omega^*$, $\eta^*$, $\{\{\mathbf{e}^*_{b,j}\}_b\}$;
    \STATE Initialize the semantic-context-aware state construction network: $\kappa$, $\tau_{j},j=\{1,2\}$;
    \STATE Initialize the resource allocation network: $\theta$, $\zeta$, $\delta_j$, $\lambda$, $\vartheta,j=\{1,2\}$;
    \FOR{Epoch $m=1$ $\rightarrow$ $M$}
    \STATE \textbf{Transmitters:}
    \STATE \par \hspace{1cm} $E_{\boldsymbol{\omega}^*}(\mathbf{S})$ $\rightarrow$ ${\boldsymbol{\varpi}}$;
    \STATE \par \hspace{1cm} $\mathbf{e}_{b,j}({\boldsymbol{\varpi}})$ $\rightarrow$ $\mathbf{e}_b$;
    \STATE Obtain the semantic-context-aware state $\tilde{s}^{(t)}$ based on \textbf{Algorithm 2};
    \STATE Obtain $a^{(t)}=\left\{\mathbf{B}^{(t)},\boldsymbol{\Psi}^{(t)},\boldsymbol{\rho}^{(t)} \right\}$ based on (32) and (37) with $\tilde{s}^{(t)}$;
    \STATE Adjust $\boldsymbol{\Psi}^{(t)}$, assign $\boldsymbol{\rho}^{(t)}$ and quantify $\mathbf{B}^{(t)}$;
    \STATE \textbf{Wireless Channel:}
    \STATE \par \hspace{1cm} Transmit indexes with OFDM;
    \STATE \textbf{Receiver:}
    \STATE \par \hspace{1cm} Recover $\{\mathbf{e}^{\prime}_{b}\}$, $D_{\eta_d}^{-1}(\mathbf{e}_{b})$ $\rightarrow$ $\mathbf{S}_d^{\prime}$;
    \STATE Compute reward $r^{(t)}$ base on (21);
    \STATE $\mathcal{T}$ $\cup$ $\left(\tilde{s}^{(t)}, a^{(t)}, \tilde{s}^{(t+1)}, r^{(t)}\right)$ $\rightarrow$ $\mathcal{T}$;
    \STATE Update $\theta$, $\zeta$, $\delta_j$, $\lambda$ and $\vartheta$ with samples from $\mathcal{T}$, online update $\kappa$ and $\tau_{j},j=\{1,2\}$;
    \ENDFOR
    \STATE Tune the whole network and update $\boldsymbol{\omega}^*$, $\boldsymbol{\eta}^*$, $\{\{\mathbf{e}^*_{b,j}\}_b\}$, $\theta$, $\zeta$, $\delta_j$, $\lambda$ and $\vartheta,j=\{1,2\}$;
    \RETURN $\boldsymbol{\omega}^*$, $\boldsymbol{\eta}^*$, $\{\{\mathbf{e}^*_{b,j}\}_b\}$, $\boldsymbol{\kappa}^*$, $\tau_{j}^*$, $\theta^*$, $\zeta^*$, $\delta^*_j$, $\lambda^*$, $\vartheta^*,j=\{1,2\}$.
  \end{algorithmic}
\end{breakablealgorithm}

\section{Simulation Results And Analyses}
This section firstly presents the simulation parameters.
Then the performance of our proposed intelligent resource allocation schemes is compared with several benchmark schemes,
including the convergence, semantic security performance and the S-SSE performance.

\subsection{Simulation Parameter Settings}
Unless specified otherwise, our simulation parameters align with those employed in \cite{9398576, 9763856, 10257607}, which are as follows.

\subsubsection{System Settings} The transmitter is located at $\left(0, 0\right)$ and is equipped with $M = 3$ antennas. 
The number of the subchannels is set as $C = 3$, each having the bandwidth $B = 90 \mathrm{~kHz}$.
Three legitimate users ($D = 3$) are respectively positioned at $\left(80, 80\right)$, $\left(100, 0\right)$ and $\left(0, 70\right)$.
The IRS is deployed at $\left(40, 40\right)$, and the number of the reflective elements is $E = 64$.
The eavesdropper equipped with one antenna is located at $\left(60, 40\right)$.
The channels from the transmitter to the IRS and from the IRS to the users follow Rician fading models, and the channels from the transmitter to the users follow Rayleigh fading models. 
The path fading is represented by $PL = \left(P L_0 - 10 \tau \log _{10}\left(d / D_0\right)\right) \mathrm{~dB}$, with $P L_0$ set as $30 \mathrm{~dB}$ and a reference distance of $D_0 = 1 \mathrm{~m}$.
The path loss exponents from the transmitter to both the target equipment and the eavesdropper, from the transmitter to the IRS, and from the IRS to both the target equipment and the eavesdropper are respectively set as $\tau_{ud} = 3$, $\tau_{ur} = 2$ and $\tau_{rd} = 2$.
The maximum transmit power of the transmitter is set as $TP = 20 \mathrm{~dBm}$.
The \textbf{Database} is set as PASCAL VOC-2007 and ImageNet.

\subsubsection{Semantic Communication Network Settings} The semantic coding network is founded upon VQ-VAE \cite{van2017neural}, 
with the learning rate of the semantic encoder and decoder set as $0.001$.
The modulation scheme is set as 64-QAM.
The size set of the codebook is $[16,32,64,128,256]$, corresponding to the maximum integer respectively represented by $[4, 5, 6, 7, 8]$ bits. 
The optimal dimension of the semantic encoder is calculated as $|{\boldsymbol{\varpi}}|=64$.
The architecture of the semantic coding network is provided in \textbf{TABLE I},
and the weighting factor $\xi$ of the human perception loss is set as $\xi=0.2$.
For the semantic-enhanced state coding networks, 
VGG-16 is employed as the foundation for the state encoders, with a learning rate of $0.001$.
The detailed framework is presented in \textbf{TABLE II}.
In NdeHDRL-DS, the Dueling-DQN incorporates full-connection networks, an advantage network and an action value network. 
SAC incorporates a policy network, two eval critic networks and two target critic networks. 
The learning rate of eval Q networks and that of the critic networks are set as $0.002$ while the learning rate of the policy network is set as $0.001$. 
The size of the memory replay is $U = 20000$. 
The other detailed settings of the algorithms in our designed semantic resource allocation networks are presented in \textbf{TABLE III}.
\begin{table*}
  \caption{Semantic Communication Network}
  \centering
  \begin{tabular}{|c|c|c|c|c|}
    \hline
    {}&Function& Layer & Units & Activation\\
    \hline
    \multirow{4}*{Transmitter}& \multirow{2}*{Semantic Encoder}&8 $\times$ Transformer &768 (12 heads)&Linear\\
    \cline{3-5}
    & & Dense &256 &ReLU\\
    \cline{2-5}
    & Pre-VQ&Conv &64& ReLU\\
    \cline{2-5}
    & Codebook&Dense &64& None\\
    \hline
    Wireless Channel& Channel& None &None &None\\
    \hline
    \multirow{5}*{Receiver}&Codebook&Dense& 64 &None\\
    \cline{2-5}
    &\multirow{2}*{Semantic Decoder}&Dense&768&ReLU\\
    \cline{3-5}
    &&8 $\times$ Transformer& 768 (12 heads)&Linear\\
    \cline{2-5}
    &Task-Attention&Dense& Task-Related Size&ReLU\\
    \hline
    \end{tabular}
\end{table*}

\begin{table}
  \caption{Semantic-Enhanced State Coding Networks}
  \centering
  \begin{tabular}{|c|c|c|}
    \hline
    & Network &Units\\
    \hline
    \multirow{5}*{VGG-16 Based State Coding}& 3 $\times$ Conv&3 $\times$ 3\\
    \cline{2-3}
    & Max-Pool & 2 $\times$ 2 \\
    \cline{2-3}
    & 4 $\times$ Conv&3 $\times$ 3\\
    \cline{2-3}
    & Max-Pool & 2 $\times$ 2 \\
    \cline{2-3}
    & Full-Connection & 1 $\times$ 64 \\
    \hline
  \end{tabular}
\end{table}

\begin{table*}
  \caption{Semantic Resource Allocation Networks}
  \centering
  \begin{tabular}{|c|c|c|c|}
    \hline
    Algorithm& Network& Hidden Neurons & Action Decision\\
    \hline
    \multirow{3}*{NoisyNet-Dueling DQN}& Full-Connection Network&3 $\times$ 512 &Bits for Semantic\\
    \cline{2-3}
    &Advantage Network& 1 $\times$ 256 & Representation \& \\
    \cline{2-3}
    &Action Value Network& 1 $\times$ 256 & Subchannel Assignment\\
    \hline 
    \multirow{2}*{NoisyNet-SAC}& Policy Network& 2 $\times$ 512 &\multirow{2}*{IRS Reflective Coefficients}\\
    \cline{2-3}
    & Critic Q Network  & 3 $\times$ 512&\\
    \hline
  \end{tabular}
\end{table*}

\subsubsection{Benchmark Scheme Settings} 
The CNN based autoencoder (\textbf{CNN-AE}) scheme \cite{yang2021deep} is used as the coding benchmark compared to our adopted \textbf{VQ-VAE} based scheme.
As two representative conventional image coding schemes, the \textbf{JPEG2000+Turbo} scheme \cite{10123081} and the \textbf{BPG+Turbo} scheme \cite{yang2022ofdm} are introduced.
the mean-square error (MSE) loss is introduced for the comparison with our designed semantic loss (10).
In order to explore the effectiveness of the proposed \textbf{SCA-SS} design and \textbf{NdeHDRL} framework,
we compare the conventional \textbf{HDRL} scheme \cite{wang2023intelligent} and the \textbf{CSO-SS} scheme for simple collocation of semantic spaces and observable state spaces.
The twin delayed deep deterministic policy gradient (TD3) is another state-of-the-art algorithm to address the continuous action space \cite{fujimoto2018addressing}.
Hence, the \textbf{NdeHDRL-DT} scheme based on NdeHDRL-dueling DQN and TD3 (NdeHDRL-DT) is also introduced for better comparison.
To further demonstrate the superior performance of our proposed DRL based intelligent resource allocation scheme,
the \textbf{Traditional Method} scheme using AO algorithm with 40 iterations \cite{10257607} is introduced.
Moreover, considering the semantic security at the symbol-related physical layer (PL-SS) proposed in \cite{wang2023star} and the semantic security at the task-related application layer (AL-SS) proposed in \cite{10123081},
the \textbf{PL-SS} scheme and the \textbf{AL-SS} scheme are compared with our proposed \textbf{CL-SS} scheme.
To verify the beneficial effects of adaptive layer selections, i.e., bits for semantic representation, with our proposed multi-layer codebook in terms of security and efficiency, 
we introduce the "Fixed Layer Selection" scheme, where the most frequent layer selected by NdeHDRL-DS is taken as the fixed selection, i.e., $\mathbf{B}$ is determined, 
and $\boldsymbol{\Psi}$ and $\boldsymbol{\rho}$ are optimized by our proposed scheme.

\subsubsection{Metric Computation} The computation of SSIM and LPIPS in eq. (2) and eq. (10) are given as follows.

\textbf{SSIM}: The mean, standard deviation and covariance are utilized respectively as estimations for brightness, contrast and structural similarity.
Let $\mu_{\mathbf{S}}$ and $\sigma_{\mathbf{S}}$ respectively be the mean value and variance of the original image $\mathbf{S}$, 
while $\mu_{\tilde{\mathbf{S}}}$ and $\sigma_{\tilde{\mathbf{S}}}$ respectively be the mean value and variance of the recovered image $\tilde{\mathbf{S}}\in \{\mathbf{S}_d^{\prime},\hat{\mathbf{S}}_e\}$,
Let $\sigma_{\mathbf{S},\tilde{\mathbf{S}}}$ be the covariance of the original image $\mathbf{S}$ and the recovered image $\tilde{\mathbf{S}}\in \{\mathbf{S}_d^{\prime},\hat{\mathbf{S}}_e\}$. 
The luminance, contrast and structure are respectively represented by
$l(\mathbf{S}, \tilde{\mathbf{S}})=\frac{2 \mu_{\mathbf{S}} \mu_{\tilde{\mathbf{S}}}+C_1}{\mu_{\mathbf{S}}^2+\mu_{\tilde{\mathbf{S}}}^2+C_1}$,
$c(\mathbf{S}, \tilde{\mathbf{S}})=\frac{2 \sigma_{\mathbf{S}} {\sigma_{\tilde{\mathbf{S}}}}+C_2}{\sigma_{\mathbf{S}}^2+\sigma_{\tilde{\mathbf{S}}}^2+C_2}$ and
$s(\mathbf{S}, \tilde{\mathbf{S}})=\frac{ \sigma_{{\mathbf{S}} {\tilde{\mathbf{S}}}}+C_3}{\sigma_{\mathbf{S}} \sigma_{\tilde{\mathbf{S}}}+C_3}$,
where $C_1$, $C_2$ and $C_3$ are given weight coefficients.
Further considering the image details at different resolutions, the multi-scale SSIM (MS-SSIM) is adopted, which can be obtained by \cite{huang2020unet}
\begin{equation}
    \operatorname{MS-SSIM}(\mathbf{S}, \tilde{\mathbf{S}}) = [l_{J}(\mathbf{S}, \tilde{\mathbf{S}})]^{\upsilon_{l,J}} \cdot \prod_{j=1}^J[c_j(\mathbf{S}, \tilde{\mathbf{S}})]^{\upsilon_{c,j}} \cdot[s_j(\mathbf{S}, \tilde{\mathbf{S}})]^{\upsilon_{s,j}},
\end{equation}
where $J-1$ stands for downsampling by a factor of 2, and we set $J = 4$ in this paper.
The $l_{j}$, $c_{j}$ and $s_{j}$ respectively represent the brightness, contrast and structural similarity at the $j$-th downsampling, while the $\upsilon_{l,j}$, $\upsilon_{c,j}$ and $\upsilon_{s,j}$ are the corresponding weighting factors.

\textbf{LPIPS:} The LPIPS is obtained by \cite{zhang2018unreasonable}
\begin{equation}
  \operatorname{LPIPS}(\mathbf{S}, \tilde{\mathbf{S}})=\sum_l \frac{1}{H_l W_l} \sum_{h, w}\left\|w_l \odot\left(\hat{y}_{h w}^l-\hat{y}_{0 h w}^l\right)\right\|_2^2,
\end{equation}
where $\hat{y}_{h w}^l$ is the unit-normalized feature stack from the $l$-th layer, and $w_l \in \mathbb{R}$ is utilized to scale the activated channels.

\vspace{-0.3cm}
\subsection{Convergence Performance}\label{s-c}
Fig. 4 demonstrates the convergence of the semantic coding network.
As shown in Fig. 4(a), we investigate the convergence rate and convergence performance.
The $0.002$ is a suboptimal learning rate compared to $0.001$ and $0.003$ when training the semantic coding in this paper.
Hence, the $0.002$ is adopted as the learning rate.
Fig. 4(b) further verifies and compares the validity of our adopted semantic coding network with several benchmark schemes versus different SNR on the validation dataset.
It is shown that the ``VQ-VAE + Semantic Loss'' scheme outperforms the ``CNN-AE + Semantic Loss'' scheme and the conventional communication schemes, ``JPEG2000/BPG + Turbo'', at a low SNR.
This is because the VQ-VAE framework can be better aware of the semantic context and effectively represent the semantics, i.e., 
encoding the data source to a prior probability distribution in the codebook-based semantic space.
Moreover, it is worth noting that the MSE loss fails to evaluate the differences at the semantic level as the ``VQ-VAE + MES Loss'' scheme 
performs poor semantic performance with severe human perception loss, especially at the low SNR, compared to the ``VQ-VAE + Semantic Loss'' scheme.
\begin{figure}
  \centering
  \setlength{\subfigcapskip}{-0.3cm}
  \subfigure[The training of our proposed VQ-VAE based semantic coding network with the training SNR = $10 \mathrm{~dB}$ and different learning rates.]{
    \begin{minipage}{10cm}
    \includegraphics[width=\textwidth]{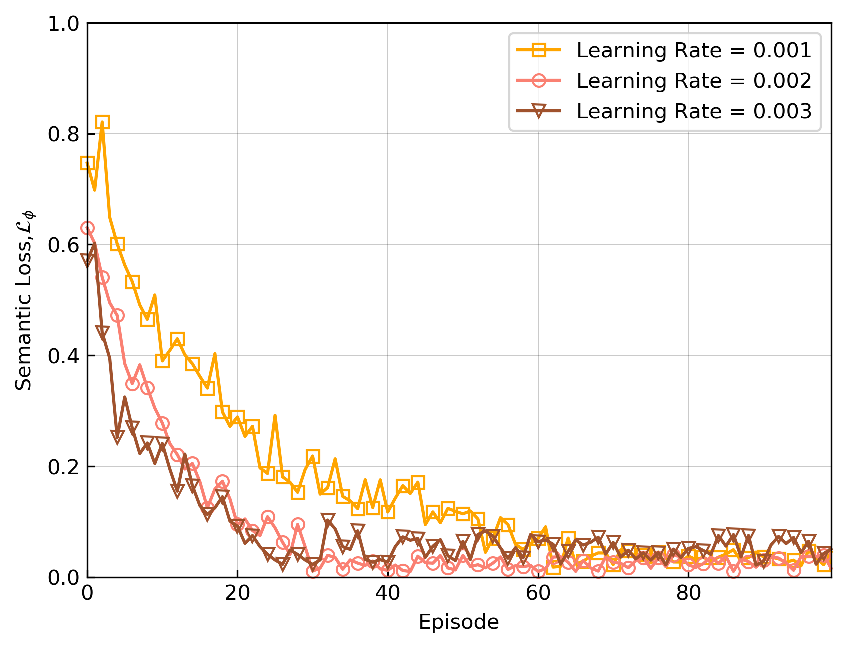} \\
    \end{minipage}}
  \subfigure[The validity of different semantic coding networks with different SNR.]{
    \begin{minipage}{10cm}
      \includegraphics[width=\textwidth]{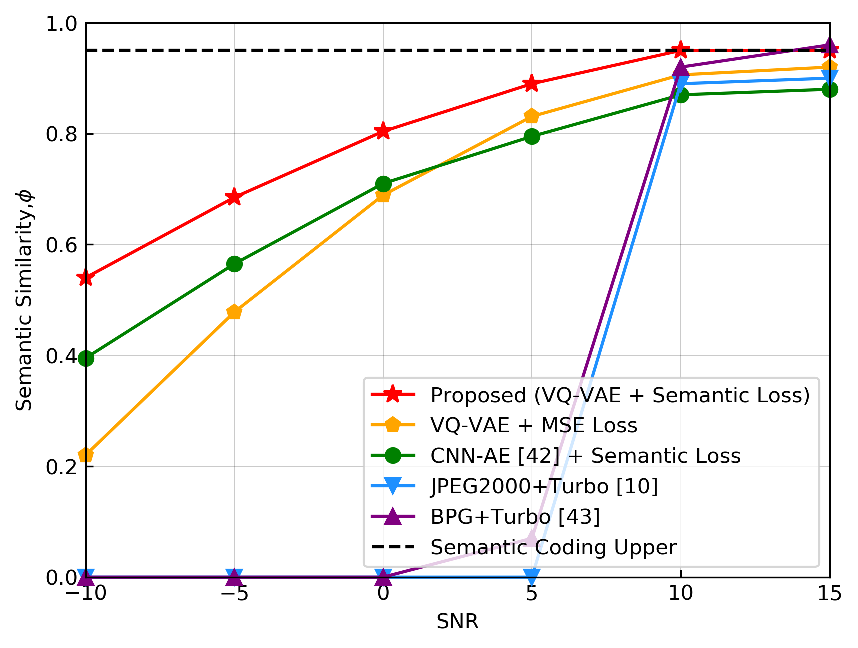} \\
    \end{minipage}}
  \caption{The convergence of our proposed semantic coding networks.} 
  \vspace{-0.5cm}
\end{figure}

Fig. 5 shows the convergence performance of our proposed intelligent resource allocation.
Our proposed NdeHDRL scheme obtains higher rewards than the HDRL scheme, 
due to the fact that the noise disturbance design can improve the generalization and stabilization capabilities of the agent.
It is also worth noting that the schemes with SCA-SS can reach the convergence in a shorter episode compared to the schemes with CSO-SS and outperform the Traditional Method scheme.
This is because the SCA-SS can effectively represent the features of the high-dimensional semantic context spaces and observable state sapces,
which mitigate the dimensionality catastrophe problem and address the sparse reward problem.
Hence, our proposed NdeHDRL-DS+SCA-SS scheme has a $42\%$ improvement of the rewards from the semantic communication environment compared to the baseline intelligent schemes.

\begin{figure}
  \centering
  \includegraphics[width=10cm]{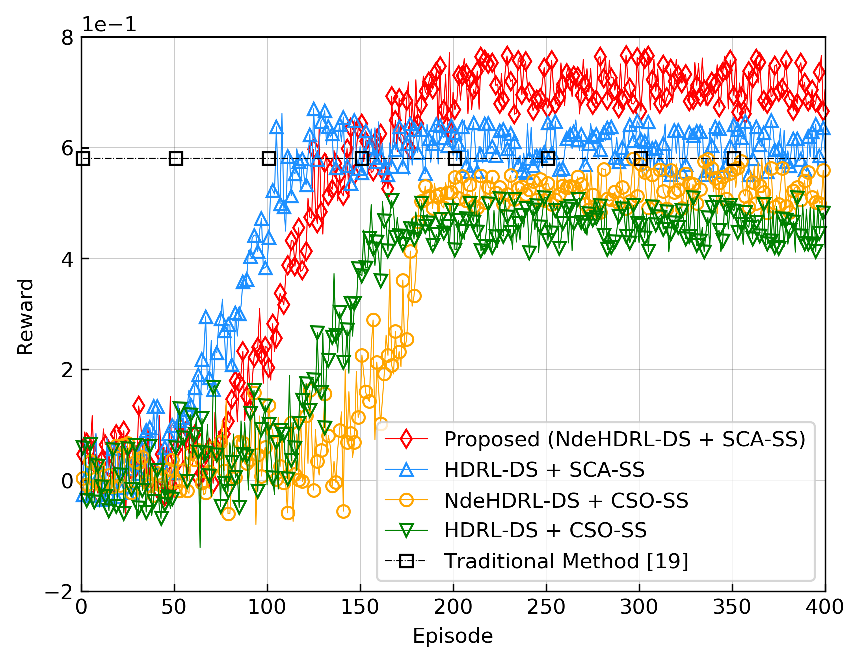}
  \vspace{-0.5cm}
  \caption{The convergence of the resource allocation schemes.}
  \vspace{-0.5cm}
\end{figure}

\subsection{Semantic Security Performance}
Fig. 6 demonstrates the semantic security performance over different levels of security.
It is shown that the PL-SS scheme is hard to ensure the semantic security with the elevated background noise. 
This is due to the fact that in the presence of the substantial background noise, a large number of error bits and semantic shifts occur at the physical layer. 
In this case, it is challenging to ensure the robust physical layer security.
Additional physical layer considerations are imperative to comprehensively address the multifaceted security requirements.
In the PL-SS scheme, the discernible semantic gap in the received messages between the target user and the eavesdropper increases with the escalation of noise intensity. 
This enhancement is attributed to the capabilities of the IRS, which improves the channel gain for the legitimate user while concurrently suppressing the eavesdropping channel.
Hence, in our proposed CL-SS scheme, we address semantic security considerations from both the semantic and physical transmission perspectives.
As shown in Fig. 6, our proposed CL-SS scheme outperforms the stand-alone PL-SS scheme and AL-SS scheme.
Moreover, it is worth noting that an upper bound exists for semantic security improvement as $\sigma$ 
increases due to the high bit error rate at the physical layer.
\begin{figure}
  \centering
  \includegraphics[width=10cm]{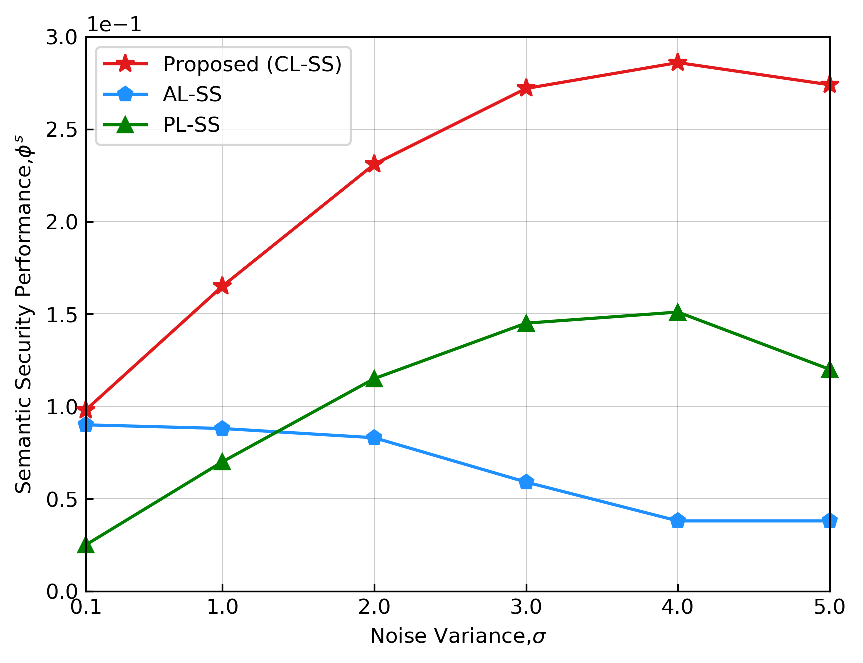}
  \vspace{-0.5cm}
  \caption{The semantic security performance of security design versus different noise variances.}
  \vspace{-0.5cm}
\end{figure}

Fig. 7 further demonstrates the semantic security performance versus different number of IRS reflective elements when $\sigma = 3$.
Both the PL-SS and CL-SS schemes achieve an increasing semantic security performance as the number of IRS reflective elements increases
since more reflective elements allow for more flexible and stable beamforming adjustment.
It demonstrates that the CL-SS scheme with NdeHDRL reaches $28.1 \%$ semantic security improvement compared to the PL-SS.
Moreover, it can be seen that our proposed NdeHDRL method can achieve $15.4 \%$ semantic security improvement compared to the HDRL-DS scheme.
This is because the robust performance of the NdeHDRL method to address the SCA-SS.
Moreover, it is difficult for the AL-SS scheme to ensure the SSC due to that there is no intervention for symbol-level protection in the physical transmission.
\begin{figure}
  \centering
  \includegraphics[width=10cm]{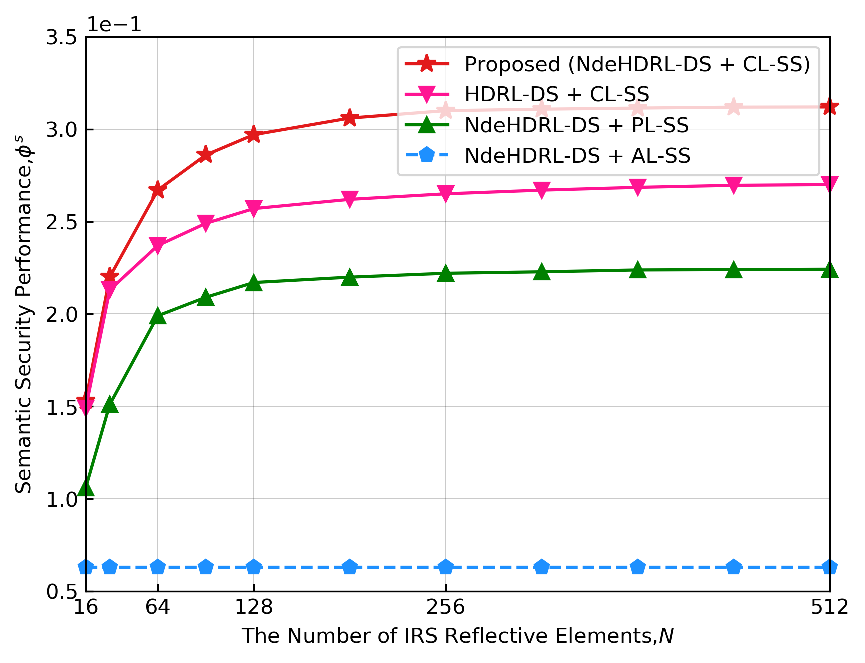}
  \vspace{-0.5cm}
  \caption{The semantic security performance versus different numbers of the IRS reflective elements.}
  \vspace{-0.5cm}
\end{figure}

\subsection{S-SSE Performance}
\begin{figure}
  \centering
  \includegraphics[width=10cm]{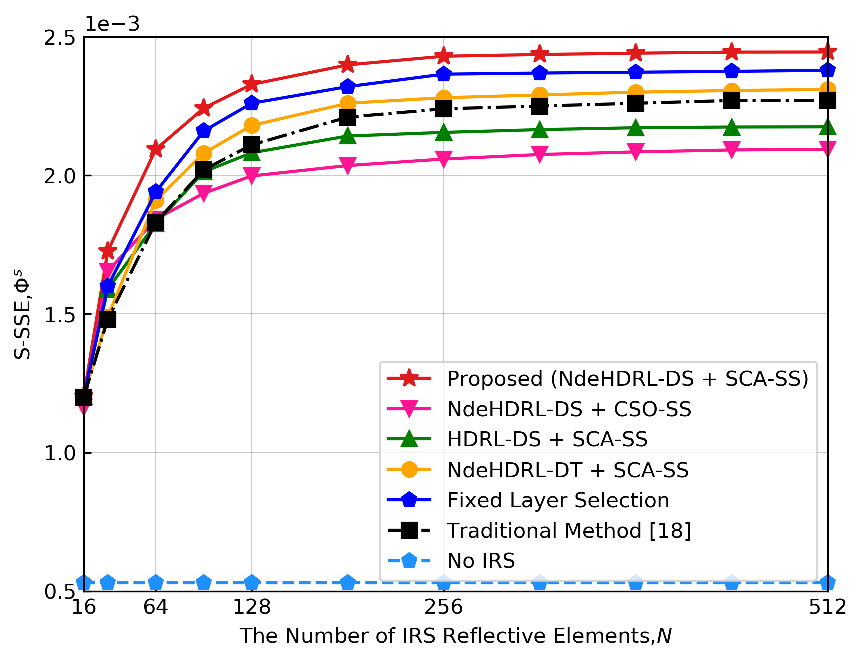}
  \vspace{-0.5cm}
  \caption{The S-SSE performance versus different numbers of the IRS reflective elements.}
  \vspace{-0.5cm}
\end{figure}
Fig. 8 investigates the effectiveness of our proposed intelligent resource allocation scheme in the S-SSE enhancement when $\sigma = 3$.
It is difficult for the scheme without the IRS to ensure the SSC,
which means that it is insufficient to consider the task-level semantic security, i.e., to train only the semantic encoding network for security, 
especially in the low signal-to-noise ratio case.
From Fig. 8, it is shown that our proposed NdeHDRL-DS scheme outperforms the NdeHDRL-DT scheme.
This is because the maximum-entropy-based SAC achieves the better exploratory capacity, which is beneficial for exploring the rich semantic context space.
Moreover, it can be seen that the increasing number of IRS reflective elements can enhance the S-SSE due to the beneficial intervention during the physical transmission stage.
From Fig. 8, when $N=128$, our proposed NdeHDRL-DS scheme with SCA-SS obtains $283 \%$ S-SSE improvement compared to the schemes without IRS,
and $16.5 \%$ S-SSE improvement compared to the NdeHDRL-DS scheme with CSO-SS.
We can also observe that our proposed scheme can outperform the Traditional Method scheme.
This is due to the fact that mathematical schemes are required to find the optimal solution in an expanded feasible solution space for semantic communication networks compared to those in conventional communication networks. 
More specifically, finding the optimal resource allocation in the physical space and finding efficient semantic representations in the semantic space are coupled in semantic communication networks. 
Hence, it is challenging for mathematical schemes to perceive the semantics and solve the expanded feasible solution space.
Compared to the "Fixed Layer Selection" scheme, 
our proposed scheme with adaptive codebook selection can ensure secure semantic communications in a more efficient manner, 
since it is tricky to fix bits for semantic representation with the mixing consideration factors including security requirements, semantic difference, task performance and resource limitations. 
In a brief summary, the efficiency of our proposed scheme can be attributed to the thorough consideration of CL-SS, well-designed SCA-SS, and robust, adaptive NdeHDRL.

\subsection{Real-Time Performance}
Analyzing computational complexity and latency for resource allocation algorithms is crucial, 
particularly in time-sensitive semantic wireless communication networks. 
Given the challenges in mathematically deriving the computational complexity of our proposed DRL-driven intelligent scheme and deep learning-driven SCA-SS, 
we assess computational efficiency and perform latency analysis by measuring the average execution time of the algorithms on central processing units (CPUs) several times. 
For better comparison, we introduce the Traditional Method scheme as a benchmark.
The simulation device is set as a four-core Intel Core i7 processor running at 2.30 GHz, with 16 GB LPDDR4X DRAM,
and the simulation result is presented in \textbf{TABLE IV}.
It is worth noting that noise perturbation enhancement for HDRL is delay-free due to its low-complexity Gaussian noise decomposition, 
and a small amount of tolerable delay is introduced during the feature extraction of observable states and semantic contexts. 
Furthermore, even with the SCA-SS design, 
our proposed scheme is over 60 times more efficient in terms of execution time compared to the Traditional Method scheme, 
demonstrating a significant improvement in real-time performance,
which allows for semantic communication networks with low-latency requirements and high intelligence.

\begin{table}
  \centering
  \caption{CPU Time Comparison of Resource Allocation Schemes for Semantic Communications}
  \begin{tabular}{|c|c|c|}
    \hline
    Scheme & Iteration & Execution Time\\
    \hline
    \multirow{3}*{Traditional Method \cite{10257607}}& 20 & 512.34 ms\\
    \cline{2-3}
    & 40 & 924.21 ms \\
    \cline{2-3}
    & 60 & 1231.67 ms\\
    \hline
    HDRL& \textbackslash & 13.67 ms\\
    \hline
    NdeHDRL& \textbackslash & 13.67 ms\\
    \hline
    NdeHDRL + SCA-SS& \textbackslash & 14.64 ms \\
    \hline
  \end{tabular}
\end{table}

\section{Conclusion}
This paper investigated a cross-layer resource allocation scheme for IRS-SSC,
where a multi-layer codebook was utilized to adaptively discretize the continuous semantic information, applicable to existing digital communication systems.
The S-SR and S-SSE were defined by mapping the security requirements of the application layer into the physical layer for the first time.
Aiming to maximize the S-SSE, the bits for semantic representations, the reflective coefficients of the IRS and subchannel assignment were jointly optimized.
The NdeHDRL scheme was considered to tackle the tough non-convex optimization problem, 
where a novel SCA-SS was able to effectively fuse the high-dimensional semantic space and observable state space.
Simulation results demonstrated that our proposed IRS-SSCs guaranteed secure semantic performance, 
and our proposed NdeHDRL scheme with CL-SS significantly enhanced the S-SSE.

\bibliography{IEEEabrv,reference}

\end{document}